\documentclass[reqno]{amsart}
\usepackage{amssymb,stmaryrd}
\usepackage{amsfonts}
\usepackage{amstext}
\usepackage{algorithmic}
\usepackage{algorithm}
\usepackage{graphicx}
\usepackage{epstopdf}
\usepackage{color}
\usepackage[all]{xy}
\usepackage{MnSymbol, slashed}

\numberwithin{equation}{section}
\newtheorem{theorem}{Theorem}[section]
\newtheorem{lemma}[theorem]{Lemma}
\newtheorem{proposition}[theorem]{Proposition}

\theoremstyle{definition}

\theoremstyle{remark}

\newcommand{\R}{{\mathbb{R}}}

\newcommand{\C}{{\mathbb{C}}}
\newcommand{\Z}{{\mathbb{Z}}}

\newcommand{\Tr}{{\rm Trace}}

\newcommand{\CL}{{\mathcal{L}}}

\newcommand{\wedgeq}{{\wedge\kern-5pt\cdot}}
\newcommand{\cg}{{\mathfrak g}}

\newcommand{\tens}{\otimes}

\newcommand{\id}{{\rm id}}

\newcommand{\extd}{{\rm d}}
\newcommand{\del}{{\partial}}
\newcommand{\eps}{\epsilon}

\renewcommand{\Tr}{{\rm Tr}}

  \allowdisplaybreaks

\begin{document}

\title{Yang-Mills fields from fuzzy sphere quantum Kaluza-Klein model }

\thanks{Ver 1.3}

\author{Chengcheng Liu and Shahn Majid}
\address{Queen Mary, University of London\\
School of Mathematics, Mile End Rd, London E1 4NS, UK}

\email{chengcheng.liu@qmul.ac.uk, s.majid@qmul.ac.uk}


\begin{abstract} Using the framework of quantum Riemannian geometry, we show that gravity on the product of spacetime and a fuzzy sphere is equivalent under minimal assumptions to gravity on spacetime, an $su_2$-valued Yang-Mills field $A_{\mu i}$ and real-symmetric-matrix valued Liouville-sigma model field $h_{ij}$ for gravity  on the fuzzy sphere. Moreover,  a massless real scalar field on the product appears as a tower of scalar fields on spacetime, with one for each internal integer spin $l$ representation of $SU(2)$, minimally coupled to $A_{\mu i}$ and with mass depending on $l$ and the fuzzy sphere size. For discrete values of the deformation parameter, the fuzzy spheres can be reduced to matrix algebras $M_{2j+1}(\C)$ for $j$ any non-negative half-integer, and in this case only integer spins $0\le l\le 2j$ appear in the multiplet. Thus, for $j=1$ a massless field on the product appears as a massless $SU(2)$ internal spin 0 field, a massive internal spin 1 field and a massive internal spin 2 field, in mass ratio $0,1,\sqrt{3}$ respectively, which we conjecture could arise in connection with an approximate $SU(2)$ flavour symmetry.  \end{abstract}
\maketitle 

\section{Introduction}\label{secintro}

The quantum spacetime hypothesis proposes that spacetime is better modelled by noncommutative coordinates due to Planck scale effects. An early work here (but not a closed spacetime algebra) was \cite{Sny} with flat spacetime models appearing in \cite{MaRue,DFR,Hoo}, some of them with quantum Poincar\'e group symmetry\cite{Luk}. Another motivation\cite{Ma:pla} was that phase space should be a quantum geometry with both momentum and position ultimately both noncommutative and curved, and in a dual relationship. To take this forward to physical applications in a non-ad hoc manner requires  coordinate-free interpretative tools and recently quantum geodesics\cite{BegMa:cur,LiuMa} have been introduced for this. This is familiar in General Relativity (GR) but applies equally well to making sense of different noncommutative coordinates even in flat models. Secondly, we need to be able to handle curved quantum spacetime in order to include gravity and quantum gravity in the models. The vision here is that quantum gravity becomes self-regularising due to Planck scale effects back on spacetime. To this end, as well as Connes' well known approach \cite{Con} coming from operator theory and cyclic cohomology, one has a more constructive `layer by layer' approach to {\em quantum Riemannian geometry} (QRG) as in \cite{BegMa} and references therein. Here, one starts with a differential structure $(\Omega^1,\extd)$ of `1-forms' on the possibly noncommutative coordinate algebra $A$, often guided by symmetry or quantum group symmetry considerations. Then a metric is $\cg\in\Omega^1\tens_A\Omega^1$ subject to certain axioms and one can ask for a `quantum Levi-Civita connection' (QLC) $\nabla:\Omega^1\to \Omega^1\tens_A\Omega^1$. This is, among other things, a bimodule connection in the sense of  \cite{DVM,Mou}. Black holes in this formalism appeared in \cite{ArgMa2} and quantum gravity models using this formalism appeared in \cite{Ma:squ,ArgMa1,LirMa1}. We refer to a recent survey \cite{ArgMa4} for an introduction and overview. 

In the present work we look at a different application to physics in which spacetime remains classical but we hope to understand the particle content of the Standard Model. The idea here first appeared using the Connes formalism  in \cite{Con1,Cham} and related works via spectral triples, but we come at it more directly using the QRG formalism. Specifically, we extend \cite{ArgMa4,LiuMa2} where quantum Kaluza-Klein theory was introduced by studying QRG on a product spacetime, where the commutative algebra of functions on usual spacetime is tensored with an internal quantum geometry $A_f$ at each point of spacetime. When $A_f=M_2(\C)$, it was shown in \cite{LiuMa2} that a real scalar field on the product appears as two scalar fields and one charged field on spacetime, with the latter coupled to a 1-form $A_\mu$. Likewise the Ricci scalar on the product appears as Ricci on spacetime, a constant contribution from Ricci on $M_2(\C)$ and a Maxwell action for $A$. In addition there were physically suppressed non-gauge invariant terms $||A.F||^2$ suggesting a deformation of the gauge theory picture of $A_\mu$ if this eventually emerges.  In this sequel, we show how nonAbelian Yang-Mills fields similarly arise very naturally from  quantum Kaluza-Klein theory by extending the previous analysis to $A_f=\C_\lambda[S^2]$, the fuzzy sphere, and this time without such unexpected terms. There is also a symmetric matrix-valued field $h_{ij}$ with sigma-model kinetic term and a matrix-Liouville-like potential in terms of $\Phi=\ln(h)$ as a matrix.  If we restrict the fuzzy sphere metric to be round then we have a single scalar field $h$ which appears as a Liouville field $\varphi=\ln(h)$. If we assume this to be a constant on spacetime then this appears as a constant parameter in the Yang-Mills action.  

The set up and initial analysis for the fuzzy sphere case first appeared in \cite{ArgMa4} but the equations for a QLC on the product appeared too complicated and were not solved, which is the main result now followed by decomposition of the resulting Ricci scalar on the product. The result then turns out to be remarkably similar for generic $\lambda\ne 0$ to the results of the Kaluza-Klein ansatz for classical $S^3$ at each point under a Kaluza-Klein ansatz, but the difference is that we are not making an ansatz. Rather, we analysed the full content of gravity on the product under minimal assumptions. Put another way, for $\lambda=0$, certain rigidities of QRG do not apply and there are many more metrics on the product, but even a small amount of noncommutativity leads us to a form of product QRG which, when $\lambda\ne 0$ singles out the Kaluza-Klein form complete with cylinder ansatz. In other words, QRG here offers an explanation of the structure of gravity and Yang-Mills as being forced by conjectured Planck-scale corrections (here, in the sphere direction) to an extended classical spacetime.  

Fuzzy spheres here have a long history, notably \cite{Str} and more recently \cite{Mad}, see also \cite{LirMa1,LirMa2,ArgMa2,BegMa}. In  these latter works and in the present paper, $\lambda$ is a continuous deformation parameter with the fuzzy sphere going to functions in $S^2$ (in an algebraic form) as $\lambda\to 0$, whereas in the physics literature `fuzzy sphere' usually means a matrix algebra thought of as `like' a sphere. Specifically, we take the {\em unit} fuzzy sphere with generators $y^i$  (we do not call them $x^i$ to avoid conflation with spacetime, which in the present paper remains classical) with relations
\begin{equation}\label{fuzzyrel} [y^i,y^j]=2\imath\lambda \eps_{ijk}y^k,\quad \sum_i (y^i)^2=1-\lambda^2,\end{equation}
where $i,j,k$ run $1,2,3$, sum over $k$ is understood and $\eps_{ijk}$ is the totally antisymmetric tensor. This is the usual angular momentum algebra (viewed as a quantum spacetime) modulo a value of its quadratic Casimir. For $\lambda =1/(2j+1)$ the spin $j$ representation of $U(su_2)$ descends to the quotient as an algebra map $\rho_j: \C_\lambda[S^2]\to M_{2j+1}(\C)$. This map is surjective but has a kernel that depends on $j$. The {\em reduced fuzzy spheres} $c_\lambda[S^2]$ are defined by quotient by this kernel, whereby 
\[\begin{array}{ccc} \C_\lambda[S^2]&{\buildrel \rho_j\over\longrightarrow}&M_{2j+1}(\C)\\ \twoheaddownarrow &\ \nearrow \cong &\\ c_\lambda[S^2]& & \end{array}\]
for such discrete values of $\lambda$. We do not limit ourselves to this discrete series, however. There is a natural differential calculus $\Omega^1(\C_\lambda[S_2])$, which restricts to the reduced case if we want and which is rotationally invariant\cite[Example~1.46]{BegMa}. More details are in the preliminaries Section~\ref{secfuzzy} but note that this has {\em three} basic 1-forms $s^i$ so that the differential structure is 3 and not 2-dimensional (this is an example of a quantum anomaly for differential structures). The  reduced case $j=1/2$ is then $M_2(\C)$ as in \cite{LiuMa2} but our results are quite different because there we took a standard 2-dimensional calculus on $M_2(\C)$. Meanwhile for generic $\lambda$, as for a classical sphere, one has
\begin{equation}\label{fuzzydec}\C_\lambda[S^2]=\oplus_{l=0}^\infty A_l\end{equation}
as used in \cite{ArgMa2,LirMa2} by decomposition as representations of $su_2$. Here the $su_2$ symmetry acts by the orbital angular momentum on the fuzzy sphere, with partial derivatives $\del_i$ conjugate to the $s^i$ acting as derivations on the fuzzy sphere by
\begin{equation}\label{deli} \del_i={1\over 2\imath\lambda }[y^i,\ ].\end{equation}
Here, $A_0$ carries in the trivial representation spanned by $1$, $A_1$ carries the spin 1-representation spanned by $\{y^i\}$ and $A_2$ carries the spin-2 representation, spanned by certain quadratic expressions. In the reduced case where $\lambda =1/j$, we have a truncation of (\ref{fuzzydec}) to
\[ M_{2j+1}(\C)=\oplus_{l=0}^{l=2j} V_l\]
where each $V_l$ is a certain subspace of matrices  carrying the spin $l$ representation via the isomorphism with $c_\lambda[S^2]$. Based on this, one can think of $\C_\lambda[S^2]$ as $M_\infty(\C)$ in some sense. We do not do this, however, as we prefer to keep the geometric picture. Also note that in previous works\cite{ArgMa2,LirMa1} the parameter $\lambda$ was the Planck scale relative to the radius of the physically scaled sphere as a quantum gravity correction, but now $\lambda$ is simply a dimensionless parameter in the unit fuzzy sphere. 

Turning to the results of the paper, after a preliminary Section~\ref{secpre} providing a recap of the formalism of QRG\cite{BegMa},   Section~\ref{secQLC} solves for a QLC on the product geometry, where the quantum metric is {\em forced}  for generic $\lambda$ to be  of the form\cite{ArgMa4}
\begin{equation}\label{prodmetric} \cg=g_{\mu\nu}\extd x^\mu\tens \extd x^\nu +   A_{\mu i}(\extd x^\mu \tens s^i+s^i\tens\extd x^\mu)+h_{ij}s^i\tens s^j,\end{equation}
with $x^\mu$  local spacetime coordinates and the coefficients fields depend only on spacetime.  We take the signature $-+++$ on $M$ and the fuzzy sphere metric $h_{ij}$ positive definite to match this spatial signature. We will be interested mainly in the round metric $h_{ij}=h\delta_{ij}$ with $h$ constant, but note that classically,  even with this assumption, there are a lot more general metrics possible and, moreover, classical $S^2$ has a 2-dimensional cotangent space so that the cross terms cannot determine three independent fields $A_{\mu i}$ (they are taken to be $su_2$-valued merely as an ansatz).

 Proceeding in the constant round metric case on the fuzzy sphere, Section~\ref{seclap} computes the Laplacian and action for a massless real scalar field $f\in C^\infty(M)\tens \C_\lambda[S^2]$ viewed on $M$ as mass $\sqrt{l(l+1)\over h}$ multiplets $\{\phi_l\}$ of different internal $SU(2)$ spin $l$.  Section~\ref{secR} computes the Ricci scalar on the product to obtain the action for gravity on the product with a constant round metric as 
\begin{align*}
S&=\int_M\prod\extd x^\mu\sqrt{-|\tilde g|}\left(\tilde R_M-\frac{3}{4h}+\frac{1}{8h}\sum_i||F^i||^2+\sum_{l=0}^\infty \phi_l^T\left(\tilde\square_{lA}-\frac{l(l+1)}{h}\right) \phi_{l}\right),
\end{align*}
where we also include a real scalar field on the product as discussed and $||\  ||$ is with respect to the effective metric $\tilde g$. Section~\ref{secgen} covers the general case where the fuzzy sphere metric $h_{ij}$, which can vary over spacetime, appears as a sigma-model field with a certain matrix-Louville-like potential as promised. 

Section~\ref{seccon} concludes with a discussion of matching with a Yang-Mills action and physical values of the coupling constant taken from, say, the weak $SU(2)$ gauge boson for purposes of comparison. Note that the Standard Model does not contain scalar fields with integer internal spin, so our model is not yet physical and the comparison is purely for illustrative purposes. Comparing with the required  weak coupling constant gives the sphere scale factor $\sqrt{h}$ as $11\lambda_P$, where $\lambda_P$ is the Planck length but the masses of the $\phi_l$ for $l>0$ are therefore out of reach. Or the fields $A_{\mu i}$ could be a hidden $SU(2)$ gauge symmetry, perhaps an unbroken version of flavour symmetry, with induced masses of order 1TeV as might be visible in a particle collider corresponding to a hypothetical very weak force comparable to that of gravity.   We also discuss some  directions for further work. So far in this approach, we have only looked as the decomposition of a scalar field on the product, whereas the Connes approach\cite{Con1,Cham} is intrinsically based on the Dirac equation. The extension of the present work to the Dirac or fermionic case with orbital-spin $l=1/2$ appearing as an internal $SU(2)$ doublet remains therefore an important next step.

\section{Preliminaries}\label{secpre}

Here, we recall some bare essentials of the formalism of quantum Riemannian geometry (QRG) as in \cite{BegMa} and references therein, as well as details of the fuzzy sphere\cite{Mad}\cite{BegMa} which we shall need. 

\subsection{Outline of QRG formalism}\label{secQRG} The formalism works over any field but for our purposes we work over $\C$ and ask that the `coordinate algebra' $A$ is a unital  $*$-algebra. Differentials are formally introduced as a bimodule $\Omega^1$ of 1-forms equipped with a map $\extd:A\to \Omega^1$ obeying the Leibniz rule 
\[ \extd(ab)=(\extd a)b+a\extd b\]
for all $a,b\in A$. This is required to extend to an exterior algebra $(\Omega,\extd)$ generated by $A,\extd A$, with $\extd^2=0$ and $\extd$ obeying the graded-Leibniz rule.  

A quantum metric is $\cg\in \Omega^1\tens_A\Omega^1$ together with a bimodule map inverse $(\ ,\ ):\Omega^1\tens_A\Omega^1\to A$ in the sense 
\begin{align}\label{invg}
( (\omega,\ )\tens\id)\cg=\omega=(\id\tens (\ ,\omega))\cg
\end{align}
for all $\omega\in \Omega^1$, and some form of quantum symmetry condition such as $\wedge(\cg)=0$ (we refer to $\cg$ as a generalised quantum metric if no form of symmetry is imposed).  The inversion condition in the classical case just says that the matrices expressing $\cg$ and $(\ ,\ )$ with respect to a basis are mutually inverse. In the quantum case it turns out, however, to force $\cg$ to be central (i.e. to commute with $a\in A$) as shown in \cite{BegMa:gra,BegMa}.  A (left) bimodule connection\cite{DVM,Mou} on $\Omega^1$ is $\nabla:\Omega^1\to \Omega^1\tens_A\Omega^1$ obeying 
\[ \nabla(a.\omega)=a.\nabla\omega+ \extd a\tens\omega,\quad \nabla(\omega.a)=(\nabla\omega).a+\sigma(\omega\tens\extd a)\]
for all $a\in A,\omega\in \Omega^1$, for some `generalised braiding' bimodule map $\sigma:\Omega^1\tens_A\Omega^1\to \Omega^1\tens_A\Omega^1$. The latter, if it exists, is uniquely determined. Classically, we would evaluate the first output of $\nabla$ against a vector field to get the associated covariant derivative, but in QRG we work directly with $\nabla$ itself. A connection is torsion free if the torsion $T_\nabla:=\wedge\nabla-\extd$ vanishes, and metric compatible if 
\[ \nabla \cg:=(\nabla\tens \id+ (\sigma\tens\id)(\id\tens\nabla))\cg\]
vanishes. When both vanish, we have a {\em quantum Levi-Civita connection} (QLC). Notice that because $\sigma$ itself depends on $\nabla$,  this is a quadratic condition, with the result that a QLC need not be unique or might not exist. The curvature of $\nabla$ is (similarly) defined as 
\[ R_\nabla=(\extd\tens \id- \id\wedge \nabla)\nabla:\Omega^1\to \Omega^2\tens_A\Omega^1.\]
Finally, working over $\C$, we need  $(\Omega,\extd)$ to be a $*$-calculus, $\cg$ `real' in the sense 
\[ \cg^\dagger=\cg;\quad \dagger:={\rm flip}(*\tens *)\]
and $\nabla$ $*$-preserving in the sense 
\begin{align}\label{creality}
 \nabla\circ *=\sigma\circ\dagger\circ\nabla,
\end{align}
see\cite{BegMa}. 

Also of interest will be the QRG Laplacian
\[ \square:=(\ ,\ )\nabla\extd: A\to A\]
which in the classical case with $\nabla$ the Levi-Civita connection recovers the Laplace-Beltrami operator. Finally, for the Ricci tensor, a  `working definition'   (but just based on copying classical formulae)  is to assume a bimodule lifting map $i:\Omega^2\to \Omega^1\tens_A\Omega^1$ and then take a trace,
\begin{align}\label{RR}
 {\rm Ricci}:=((\ ,\ )\tens \id)(\id\tens i\tens\id)(\id\tens R_\nabla)\cg\in \Omega^1\tens_A\Omega^1,\quad R_A:=(\ , ){\rm Ricci}
\end{align}
where $R_A$ is then the Ricci scalar curvature. Note that the classical cases of ${\rm Ricci}$ and $R_A$ in the natural conventions here are  $-{1/2}$ of their usual values\cite{BegMa}.

\subsection{Reference QRG on fuzzy sphere}\label{secfuzzy}

Here $A= \C_\lambda[S^2]$ is the algebra with generators $y^i$ and the relations (\ref{fuzzyrel}). For $\Omega$ we take this as freely generated with a central 1-form basis $s^i$, $i=1,2,3$ and exterior derivative obeying\cite[Example~1.46]{BegMa},
\[ \extd y^i= \eps_{ijk} y^j s^k,\quad [y^i,s^j]=0,\quad \extd s^i=-{1\over 2}\eps_{ijk}s^j\wedge s^k,\quad \{s^i,s^j\}=0.\]
From the Leibniz rule one can deduce that the associated partial dertivatives $\del_i$, defined by $\extd f=(\del_i f)s^i$ for any $f\in A$, are given by (\ref{deli}), from which it is clear that they act on the vector space spanned by $\{y^i\}$ in the orbital spin 1 representation and on products as the tensor product representation. For example, the elements
\[ y^iy^j+y^j y^i -{2(1-\lambda ^2)\over 3}\delta_{ij} \]
(only five of them are linearly independent) are for the form $f_{ij}y^iy^j$ where $f_{ij}$ is symmetric and traceless, hence transform in the orbital spin 2 representation. These orbital spins on the internal fuzzy sphere will appear as an internal $SU(2)$ spin of functions on the product when viewed as fields on spacetime. The geometric meaning of the $s^i$ is similarly that they correspond classically to the 1-forms associated via the metric to the killing vector fields for the $su_2$ action. Here, one has a formula\cite{LirMa2}
\begin{equation}\label{si} s^i=\frac{1}{1-\lambda^2}\left(\frac{1}{2\imath\lambda }y^i y^j\extd y^j+\epsilon_{ijm}(\extd y^j)y^m\right)\end{equation}
for the $s^i$ as differential forms, where on a classical sphere, $y^i\extd y^i=0$ and we would not have the first term. 

Next, a QLC is known for any quantum metric $\cg=h_{ij}s^i\tens s^j$ for a real symmetric matrix $h_{ij}$ (but we will be interested mainly in the `round metric' where $h_{ij}=h\delta_{ij}$ with an overall scale factor $h$). The QLC from \cite{LirMa2,ArgMa4} is
\[ \nabla s^i=H^i{}_{jk}s^j\tens s^k;\quad H^i{}_{jk}=-{1\over 2}h^{im}(2\eps_{mkl}h_{lj}+\Tr(h)\eps_{mjk}),\quad \sigma(s^i\tens s^j)=s^j\tens s^i.\]
For the standard anti-symmetric lift $i(s^i\wedge s^j)=(s^i\tens s^j-s^j \tens s^i)/2$, the Ricci scalar comes out as 
\begin{equation}\label{Rfuzzy} R_A={1\over 2\det(h)}\left(\Tr(h^2)-{1\over 2}\Tr(h)^2\right).\end{equation}
We also note the QRG Laplacian
\begin{equation}\label{lapfuzzy} \square_A=(\ ,\ )\nabla\extd=h^{ij}\del_i\del_j+ h^{kl}H^i{}_{kl}\del_i=h^{ij}\del_i\del_j,\end{equation}
where $h^{ij}$ is the inverse matrix to $h_{ij}$ and where the second term drops out by symmetry arguments for stated form of the QLC.

We will be mostly concerned with the special case of the {\em round metric} $h_{ij}=h\delta_{ij}$ on the fuzzy, in which case these become
\[ \nabla s^i=-{1\over 2} \eps_{ijk}s^j\tens s^j,\quad H^i{}_{jk}=-{1\over 2}\eps_{ijk},\quad R_A=-{3\over 4h},\quad \square_A=\sum_i \frac{1}{h}\del_i^2.\]
In that case,  $\square_A$ is diagonal on the subspaces $A_l$ with 
\begin{equation}\label{squareAl} \square_A|_{A_l}=-\frac{l(l+1)}{h}\id \end{equation}
as is clear from the representation theory or by direct calculation. Explicitly, if we expand we can write any $f\in A_l$ as 
\[f= f_{i_1...i_l}y^{i_1}y^{i_2}...y^{i_l}\]
for a corresponding  totally symmetric and traceless tensor $f_{i_1...i_l}$, and compute
\begin{align}\label{fyyy}
\del_i(f_{i_1...i_l}y^{i_1}y^{i_2}...y^{i_l})=-(\epsilon_{kii_1}f_{i_2i_3...i_lk}+\epsilon_{kii_2}f_{i_3...i_li_1k}+...+\epsilon_{kii_l}f_{i_1...i_{(l-1)}k})y^{i_1}y^{i_2}...y^{i_l}
\end{align}
Then one can directly check that
\begin{align}\label{doubledel}
\partial_i \partial_i (f_{i_1...i_l}y^{i_1}y^{i_2}...y^{i_l})=-l(l+1)f_{i_1...i_l}y^{i_1}y^{i_2}...y^{i_l}
\end{align}
where we sum over $i$.  

In the model, we will define $su_2$ Lie algebra generators $T^i=\imath\del_i$ acting on $A_l$. Thinking of the elements of $A_l$ equivalently via their corresponding totally symmetric traceless tensors $f_{i_1\cdots i_l}$, the action on the space of such tensors is then 
\[(T^i f)_{i_1...i_l}=-\imath(\epsilon_{kii_1}f_{i_2i_3...i_lk}+\epsilon_{kii_2}f_{i_3...i_li_1k}+...+\epsilon_{kii_l}f_{i_1...i_{(l-1)}k})\]
which is also totally symmetric and traceless for indices $i_1,...,i_l$. 

Finally, there is a notion of integration $\int: \C_\lambda[S^2]\to \C$ given in \cite{LirMa2} as follows. We take any $f\in \C_\lambda[S^2]$ and decompose it as
\[ f=f_01+ f_i y^i+ f_{ij}y^iy^j+\cdots+f_{i_1...i_l}y^{i_1}y^{i_2}...y^{i_l}+\cdots\]
and then $\int f=f_0$. From this one can show, for example, that\cite{ArgMa2},
\[ \int 1=1,\quad\int y^i=0,\quad \int y^iy^j=
{1-\lambda ^2\over 3}\delta^{ij}.\]
The latter can also be deduced since it must be a multiple of $\delta^{ij}$ with the multiple fixed by the integral in degree 0 applied to $\delta_{ij}\int y^iy^j$. Similarly, one has 
\[ \int y^iy^jy^k={\imath\lambda(1-\lambda^2)\over 3}\eps^{ijk},\quad \int y^iy^jy^ky^l={1-\lambda^2\over 15}\left((1+\lambda^2)(\delta^{ij}\delta^{kl}+\delta^{il}\delta^{jk})+(1-9\lambda^2)\delta^{ik}\delta^{jl} \right).\]
Rotationally invariant tensors must be built from $\delta,\eps$  and hence the only possibility in degree 3 is a multiple of $\eps^{ijk}$, with the multiple determined from $\int y^i[y^j,y^k]$ and the integral in degree 2. For degree 4, the most general tensor built from $\delta,\eps$ has the form
\[\int y^iy^jy^ky^l=\alpha \delta^{ij}\delta^{kl}+\beta\delta^{ik}\delta^{jl}+\gamma\delta^{il}\delta^{jk} \]
 but has to  be invariant under cyclic rotation of the indices as $\int$ is known to be a trace (so $\int y^iy^jy^ky^l=\int y^ly^iy^jy^k$). This and the integral  in degree 3 applied to $\int y^iy^j[y^k,y^l]$ and the integral in degree 2 applied to  $\delta_{ij}\int y^iy^jy^ky^l$ give
\[ \gamma=\alpha,\quad {2\lambda^2(1-\lambda^2)\over 3}=\alpha-\beta,\quad {(1-\lambda^2)^2\over 3}=4\alpha+\beta.\]
with solution as stated.

\section{Quantum Levi-Civita connection on $C^\infty(M)\tens \C_\lambda[S^2]$}\label{secQLC}

The coordinate algebras $C^\infty(M)$ and $\C_\lambda[S^2]$ mutually commute in the tensor product algebra and we specify the product differential calculus by requiring that this applies also in a graded sense for the exterior algebra (that this is a graded tensor product of the exterior algebra on each algebra). In particular,  1-forms on one commute with elements of the other.
 
We recall for generic $\lambda$ that the most general quantum metric on the tensor product algebra $C^\infty(M)\tens \C_\lambda[S^2]$ then has the form\cite{ArgMa4}  
\begin{align}
\cg=g_{\mu\nu}(x,t)\extd x^\mu\tens \extd x^\nu+A_{i\mu}(x,t)(\extd x^\mu\tens s^i+s^i\tens \extd x^\mu)+h_{ij}(x,t)s^i\tens s^j
\end{align}
where $g_{\mu\nu},A_{i\mu},h_{ij}$ are symmetric. There is no $\C_\lambda[S^2]$ dependence of the coefficients as the metric has to be central, the basis is central and the centre of the algebra is trivial provided we exclude $\lambda^2=1$.

We define the inverse of the metric as
\[\tilde g^{\mu\nu}=(\extd x^\mu,\extd x^\nu),\quad \tilde h^{ij}=(s^i,s^j),\quad \tilde A^{i\mu}=\tilde A^{\mu i}=(\extd x^\mu,s^i)=(s^i,\extd x^\mu),\]
which to be inverse to (\ref{invg}) requires the group of four conditions
\begin{equation*}
\begin{gathered}
g_{\mu \nu} \tilde{A}^{i \nu}+A_{j \mu} \tilde{h}^{i j}=0, \quad g_{\mu \gamma} \tilde{g}^{\gamma \nu}+A_{i \mu} \tilde{A}^{i \nu}=\delta_\mu^\nu\\
A_{i \mu} \tilde{g}^{\mu \nu}+h_{i j} \tilde{A}^{j \nu}=0 , \quad h_{i k} \tilde{h}^{k j}+A_{i \mu} \tilde{A}^{j \mu}=\delta_i^j. 
\end{gathered}
\end{equation*}
The third condition gives
\begin{align}\label{Atilde}
\tilde{A}^{i \mu}=-\tilde{g}^{\mu \nu}h^{i j}A_{j \nu},
\end{align}
then the second condition gives
\begin{align}\label{gtilde}
\tilde g_{\mu\nu}=g_{\mu\nu}-h^{ij}A_{i\mu}A_{j\nu}
\end{align}
and the first and fourth conditions give
\begin{align}
\tilde h_{ij}=h_{ij}-g^{\mu\nu}A_{i\mu}A_{j\nu}. 
\end{align}
Then (\ref{Atilde}) and the fourth condition give
\begin{align}\label{hgequation}
\tilde h^{ij}=h^{ij}+h^{ik}h^{lj}\tilde g^{\mu\nu}A_{k\nu}A_{l\mu}. 
\end{align}

Now we consider the most general form of connection,
\begin{align*}
\nabla \extd x^\mu=-\Gamma^\mu_{\alpha\beta}\extd x^\alpha\tens \extd x^\beta+F^\mu_{\alpha i}\extd x^\alpha\tens s^i+F^\mu_{i \alpha} s^i\tens \extd x^\alpha+D^\mu_{ij} s^i\tens s^j,\\
\nabla s^k=E^k_{\alpha\beta}\extd x^\alpha\tens \extd x^\beta+B^k_{\alpha i}\extd x^\alpha\tens s^i+B^k_{i\alpha } s^i\tens \extd x^\alpha+H^k_{ij} s^i\tens s^j.
\end{align*}
Requiring the torsion free condition  $\wedge\nabla-\extd=0$, we get 
\begin{align*}
\Gamma_{\alpha \beta}^\mu=\Gamma_{\beta \alpha}^\mu,\quad F^\mu_{\alpha i}=F^\mu_{i\alpha} \quad D_{k l}^\mu=D_{l k}^\mu, \quad E_{\alpha \beta}^i=E_{\beta \alpha}^i,\quad B^k_{\alpha i}=B^k_{i\alpha}, \quad H_{j k}^i-H_{k j}^i+\epsilon_{i j k}=0.
\end{align*}
We assume that these coefficients have a classical limit, i.e. do not blow up as $\lambda\to 0$ and we require that $\nabla$ is a bimodule connection.

\begin{lemma} For $\nabla$ to be a bimodule connection for $\lambda$ generic:
\begin{enumerate}\item[(i)] The connection coefficients must be at most linear in the $y^i$.
\item[(ii)] The coefficients of $\sigma$ in the combined basis $\{\extd x^\mu,s^i\}$ must depend only on spacetime.
\item[(iii)]  $\sigma(\extd x^\mu\tens \extd x^\nu)=\extd x^\nu\tens\extd x^\mu$, $\sigma(s^i\tens\extd x^\mu)=\extd x^\mu\tens s^i$, and the other two cases are the flip map plus  $O(\lambda)$ corrections. 
\item[(iv)] If these corrections vanish (i.e. $\sigma$ is the flip on all the combined basis) then the connection coefficients must depend only on spacetime.
\end{enumerate}
\end{lemma} 
\proof  Using $\nabla(e^\alpha a)=\nabla(a e^\alpha)$ for $e^\alpha$ denoting either $\extd x^\mu$ or $s^i$, one can deduce
\[\sigma(e^\alpha\tens\extd a)=[a,\nabla e^\alpha]+\extd a\tens e^\alpha\]
for all $a\in C^\infty(M)\tens \C_\lambda[S^2]$ from which it follows that
\begin{align*}
\sigma(e^\alpha\tens\extd x^\mu)=[x^\mu,\nabla e^\alpha]+\extd x^\mu\tens e^\alpha=\extd x^\mu\tens e^\alpha
\end{align*}
and
\begin{align*}
&\sigma(e^\alpha\tens s^i)=\sigma\left(e^\alpha\tens \frac{1}{1-\lambda^2}\left(\frac{1}{2\imath\lambda }y^i y^j\extd y^j+\epsilon_{ijm}(\extd y^j)y^m\right)\right)\\
&=\frac{1}{2\imath\lambda(1-\lambda^2)}y^i y^j\sigma\left(e^\alpha\tens \extd y^j\right)+\frac{1}{1-\lambda^2}\epsilon_{ijm}\sigma\left(e^\alpha\tens\extd y^j\right)y^m\\
&=\frac{1}{2\imath\lambda(1-\lambda^2)}y^i y^j([y^j,\nabla e^\alpha]+\extd y^j\tens e^\alpha)+\frac{1}{1-\lambda^2}\epsilon_{ijm}([y^j,\nabla e^\alpha]+\extd y^j\tens e^\alpha)y^m\\
&=\frac{1}{2\imath\lambda(1-\lambda^2)}\left(y^i y^j[y^j,\nabla e^\alpha]+2\imath\lambda\epsilon_{ijm}[y^j,\nabla e^\alpha] y^m\right)+s^i\tens e^\alpha,
\end{align*}
where in the first and last steps we used the expression (\ref{si}) for $s^i$. Next, writing $\nabla e^\alpha=K^\alpha_{\beta\gamma}e^\beta\tens e^\gamma$ for our merged basis, 
\begin{align}\label{sigma2}
\sigma(e^\alpha\tens s^i)=\frac{1}{2\imath\lambda(1-\lambda^2)}\left(y^i y^j[y^j,K^\alpha_{\beta\gamma}]+2\imath\lambda\epsilon_{ijm}[y^j,K^\alpha_{\beta\gamma}] y^m\right)e^\beta\tens e^\gamma+s^i\tens e^\alpha
\end{align}
We also note that $\sigma$ being a bimodule map includes
\begin{align}\label{bimodule}
a\sigma(e^\alpha\tens e^\beta)=\sigma(e^\alpha\tens e^\beta) a
\end{align}
for all $a\in C^\infty(M)\tens \C_\lambda[S^2]$, so $\sigma(e^\alpha\tens e^\beta)$ commutes with $\C_\lambda[S^2]$. Here $\sigma(e^\alpha\tens\extd x^\mu)$ satisfies (\ref{bimodule}) automatically, but for $\sigma(e^\alpha\tens s^i)$, this implies for generic $\lambda$ that
\begin{align}\label{KKK}
y^i y^j[y^j,K]+2\imath\lambda\epsilon_{ijm}[y^j,K] y^m=c_i,
\end{align}
where $c_i$ are constant in $\C_\lambda[S^2]$. For simplicity, we denote $K^\alpha_{\beta\gamma}$ by $K$ as well as $c_i^\alpha{}_{\beta\gamma}$ by $c_i$, and will recover the labels later.

 Next we multiply $y^i$ on both sides of (\ref{KKK}) and sum, giving
\begin{align}\label{K2}
(1-\lambda^2)y^j[y^j,K]+2\imath\lambda\epsilon_{ijm}y^i[y^j,K] y^m=c_i y^i
\end{align}
To solve above equation, we  note that since
\[0=[y^j y^j,K]=y^j[y^j,K]+[y^j,K]y^j,\]
we have $y^j[y^j,K]=-[y^j,K]y^j$. Then the first term on the left hand side of (\ref{K2}) can be written as
\begin{align*}
(1-\lambda^2)y^j[y^j,K]=&\frac{1}{2}(1-\lambda^2)(y^j[y^j,K]-[y^j,K]y^j)=\frac{1}{2}(1-\lambda^2)[y^j,[y^j,K]]\\
=&2\lambda^2(\lambda^2-1)\del_j\del_j K=2\lambda^2(\lambda^2-1)\del_j\del_j (K_0+\sum_{l=1}^\infty K_{i_1...i_l}y^{i_1}y^{i_2}...y^{i_l})\\
=&2\lambda^2(1-\lambda^2)\sum_{l=1}^\infty l(l+1)K_{i_1...i_l}y^{i_1}y^{i_2}...y^{i_l},
\end{align*}
where we use (\ref{deli}) in the third step, expand $K$ in the fourth step into modes of different $l$, and use (\ref{doubledel}) in the fifth step. Now we calculate the second term on the left hand side of (\ref{K2}),
\begin{align*}
2\imath \lambda\epsilon_{ijm}y^i[y^j,K] y^m
=&2\imath \lambda\epsilon_{ijm}y^iy^jK y^m-2\imath \lambda\epsilon_{ijm}y^iKy^j y^m\\
=&\imath \lambda\epsilon_{ijm}[y^i,y^j]K y^m-\imath \lambda\epsilon_{ijm}y^iK[y^j, y^m]\\
=&\imath \lambda\epsilon_{ijm}(2\imath\lambda\epsilon_{ijk}y^k)K y^m-\imath \lambda\epsilon_{ijm}y^iK(2\imath\lambda\epsilon_{jmk}y^k)\\
=&-4\lambda^2 y^k K y^k+4\lambda^2 y^k K y^k=0.
\end{align*}
As a result, (\ref{K2}) reduces to
\begin{align*}
2\lambda^2(1-\lambda^2)\sum_{l=1}^\infty l(l+1)K_{i_1...i_l}y^{i_1}y^{i_2}...y^{i_l}=c_i y^i
\end{align*}
giving
\begin{align}\label{K111}
K_{i}=\frac{c_i}{4\lambda^2(1-\lambda^2)},\quad K_{i_1...i_l}=0,\quad l\geq 2
\end{align}
Thus we obtain
\begin{align}\label{Ksolve}
K=K_0+\frac{c_i}{4\lambda^2(1-\lambda^2)}y^i
\end{align}
One can show that (\ref{Ksolve}) also satisfies (\ref{KKK}), so (\ref{Ksolve}) is the most general solution of (\ref{KKK}). 

Now substituting (\ref{KKK}),(\ref{K111}) back into (\ref{sigma2}), we obtain
\[
\sigma(e^\alpha\tens s^i)=\frac{c_i}{2\imath\lambda(1-\lambda^2)}e^\beta\tens e^\gamma+s^i\tens e^\alpha=\frac{4\lambda^2(1-\lambda^2)K_i}{2\imath\lambda(1-\lambda^2)}e^\beta\tens e^\gamma+s^i\tens e^\alpha\]
in our notation. Putting back the suppressed indices, we have 
\[ \sigma(e^\alpha\tens s^i)=-2\imath \lambda K^\alpha_{i\beta\gamma}e^\beta\tens e^\gamma+s^i\tens e^\alpha,\]
which we see deforms the flip map. The same then applies to $\sigma(s^i\tens e^\alpha)$ since $\sigma(s^i\tens s^j)$ is already covered by the above and $\sigma(s^i\tens\extd x^\mu)=\extd x^\mu\tens s^i$ as we already showed.  

For the last part, if we assume $\sigma(e^\alpha\tens s^i)=s^i\tens e^\alpha$, i.e. $K^\alpha_{i\beta\gamma}=0$, then (\ref{Ksolve}) becomes that $K^\alpha_{\beta\gamma}$ has only the $K_0$ component constant in $y^i$.   \endproof

In view of this lemma, we henceforth make the further simplifying assumption  that $\sigma$ between any basis 1-forms $s^i, \extd x^\mu$ is  a flip. The lemma says this is already true for two cases and we assume the possible $O(\lambda)$ corrections in the other two cases are not present, i.e. we assume
\begin{equation}\label{sigassump} \sigma(s^i\tens  s^j)=s^j\tens s^i,\quad \sigma(\extd x^\mu\tens s^i)=s^i\tens\extd x^\mu.\end{equation}
The first of these is usually taken for the fuzzy sphere quantum geometry and assuming it will ensure that we land there in that sector of the QRG. Then by the lemma, the connection coefficients, like the metric, depend only on spacetime, and one can check that we then indeed have a bimodule connection.

Next,  the seven equations for $\nabla\cg=0$ in \cite{ArgMa2}, which we do not need in full (so we omit them) reduce to a group of six equations:
\begin{align*}
& \partial_\alpha g_{\beta \gamma}-g_{\mu \gamma} \Gamma_{\alpha \beta}^\mu-g_{\beta \mu} \Gamma_{\alpha \gamma}^\mu+A_{i \gamma} E_{\alpha \beta}^i+A_{i \beta} E_{\alpha \gamma}^i=0, \\
& \nabla_\alpha A_{i \beta}+g_{\beta \mu} F_{\alpha i}^\mu+A_{k \beta} B_{\alpha i}^k+h_{k i} E_{\alpha \beta}^k=0, \\
& g_{\mu \beta} F_{\alpha i}^\mu+g_{\alpha \mu} F_{\beta i}^\mu+A_{k \beta} B_{\alpha i}^k+A_{k \alpha} B_{\beta i}^k=0, \\
& \partial_\alpha h_{i j}+A_{i \mu} F_{\alpha j}^\mu+A_{j \mu} F_{\alpha i}^\mu+h_{k j} B_{\alpha i}^k+h_{i k} B_{\alpha j}^k=0, \\
& g_{\mu \alpha} D_{i j}^\mu+A_{j \mu} F_{\alpha i}^\mu+A_{k \alpha} H_{i j}^k+h_{j k} B_{\alpha i}^k=0, \\
& A_{j \mu} D_{i k}^\mu +A_{k \mu} D_{i j}^\mu+h_{l k} H_{i j}^l+h_{l j} H_{i k}^l=0,
\end{align*}
where $\nabla$ is the covariant derivative defined by $\Gamma^\mu_{\alpha\beta}$. To solve these equations, we first consider the symmetric part of the second equation in our group,
\[\nabla_{(\alpha} A_{ \beta)i}+g_{\mu(\beta } F_{\alpha) i}^\mu+A_{k (\beta} B_{\alpha) i}^k+h_{k i} E_{(\alpha \beta)}^k=0,\]
minus the third equation, to obtain
\begin{align}\label{sE}
E_{\alpha \beta}^k=-\frac{1}{2}h^{ki}\nabla_{(\alpha} A_{ \beta)i}. 
\end{align}
Substituting (\ref{sE}) into the second equation in our group, we have
\begin{align}\label{sB1}
F^\mu_{\alpha i}=-g^{\beta\mu}(\frac{1}{2}\del_{[\alpha} A_{ \beta]i}+A_{k\beta}B^k_{\alpha i}), 
\end{align}
where we replace $\nabla_\alpha$ by $\del_\alpha$ as it appears antisymmetrised, due to the connection being torsion free.

Next we consider the antisymmetric part of fifth equation in our group, 
\[F_{\alpha [i}^\mu A_{j] \mu}-A_{k \alpha}\epsilon_{ijk}+ B_{\alpha [i}^k h_{j] k}=0\]
plus the fourth equation, giving
\begin{align}
B^k_{\alpha i}=-h^{jk}A_{j\mu}F^\mu_{\alpha i}+\frac{1}{2}h^{jk}(A_{l\alpha}\epsilon_{ijl}-\del_\alpha h_{ij}). 
\end{align}
Combining above equation with (\ref{sB1}), we obtain $B,F$ as:
\begin{align}
F^\mu_{\alpha i}=\frac{1}{2}\tilde g^{\beta\mu}(-\del_{[\alpha}A_{\beta]i}+h^{jk}A_{k\beta}(\del_\alpha h_{ij}-A_{\alpha l}\epsilon_{ijl})),\label{F2}\\
B^k_{\alpha i}=\frac{1}{2}\tilde h^{jk}(g^{\mu\beta}A_{j\mu}\del_{[\alpha}A_{\beta]i}+A_{l\alpha}\epsilon_{ijl}-\del_\alpha h_{ij}).\label{B2}
\end{align}

Next, we consider the symmetric part of the fifth equation in our group,
\[g_{\mu \alpha} D_{(i j)}^\mu+ F_{\alpha (i}^\mu A_{j) \mu}+A_{k \alpha} H_{(i j)}^k+ B_{\alpha (i}^k h_{j) k}=0\]
minus
fourth equation, giving
\begin{align}\label{sD1}
D^\mu_{ij}=\frac{1}{2}g^{\mu\alpha}(\partial_\alpha h_{i j}-A_{k\alpha}H^k_{(ij)}).
\end{align}
We now consider the antisymmetric part of the sixth equation,
\[A_{j \mu} D_{[i k]}^\mu +A_{\mu [k } D_{i] j}^\mu+h_{l [k} H_{i] j}^l+h_{l j} H_{[i k]}^l=0\]
which, in using zero torsion and relabelling indices, is 
\[A_{\mu [j } D_{k] i}^\mu+h_{l [j} H_{k] i}^l-h_{l i}\epsilon_{lkj}=0.\]
Adding the above equation to the sixth equation in our group gives
\[2A_{j \mu} D_{i k}^\mu+h_{kl}H^l_{[ij]}+h_{jl} H^l_{(ik)}-h_{l i}\epsilon_{lkj}=0\]
or
\[2A_{j \mu} D_{i k}^\mu+h_{jl} H^l_{(ik)}=h_{kl}\epsilon_{lij}+h_{l i}\epsilon_{lkj}.\]
Now we combine the above equation with (\ref{sD1}), to give
\begin{align}\label{sH1}
H^l_{(ik)}=\tilde h^{jl}(h_{n(i}\epsilon_{k)jn}-A_{j\mu}g^{\mu\alpha}\del_\alpha h_{ik})
\end{align}
and to obtain $D$ as
\begin{align}\label{sD}
D^\mu_{ij}=\frac{1}{2}\tilde g^{\mu\alpha}(\del_\alpha h_{ij}-A_{k\alpha}h^{mk}h_{n(i}\epsilon_{j)mn}). 
\end{align}
Combining (\ref{sH1}) with $H^l_{[ij]}=-\epsilon_{lij}$, we obtain $H$ as
\begin{align}\label{sH}
H^l_{ik}=\frac{1}{2}\tilde h^{jl}(h_{n(i}\epsilon_{k)jn}-A_{j\mu}g^{\mu\alpha}\del_\alpha h_{ik})-\frac{1}{2}\epsilon_{lik}. 
\end{align}
Finally, one can check that to satisfy the first equation, we obtain
\begin{align}\label{G2}
\Gamma^\sigma_{\mu\nu}=\tilde\Gamma^\sigma_{\mu\nu}+\frac{1}{2}\tilde g^{\sigma\rho}(A_{\mu i} \del_{[\nu} (A_{\rho]j}h^{ij})+A_{\nu i} \del_{[\mu} (A_{\rho]j}h^{ij})+A_{i\mu}A_{j\nu}\del_\rho h^{ij}),
\end{align}
again replacing $\nabla$  by $\del$ as it enters antisymmetrised. Here,  $\tilde\Gamma^\sigma_{\mu\nu}$ are the classical Levi-Civita connection coefficients for the effective metric $\tilde g_{\mu\nu}$. Looking at our results and noting that $\tilde g_{\mu\nu}$ was also given in (\ref{gtilde}) in terms of $g_{\mu\nu},h_{ij},A_{\mu i}$, we have proven:

\begin{theorem} For any quantum metric $\cg$ as in (\ref{prodmetric}) on the product, there is a unique QLC given by the solutions obtained above.
\end{theorem}

For the next two sections, we will focus on the round metric on the fuzzy sphere of constant radius, i.e. 
\begin{equation}\label{round} h_{ij}=h\delta_{ij},\quad \del_\alpha h=0.\end{equation}
Then our results simplify to
 \begin{align}
& F^\mu_{\beta i}=\frac{1}{2}\tilde g^{\alpha \mu}(\del_{[\alpha}A_{\beta]i}-\frac{1}{h}A_{\alpha j}A_{\beta k}\epsilon_{ijk}),\\ 
&\Gamma^\sigma_{\mu\nu}=\tilde\Gamma^\sigma_{\mu\nu}-\frac{1}{h}(A_{\mu i}F^\sigma_{\nu i}+A_{\nu i}F^\sigma_{\mu i}),\\
&D^\mu_{ij}=0,\\
&E^k_{\alpha\beta}=-\frac{1}{2h}\nabla_{(\alpha}A_{\beta)k},\\
&B^k_{\alpha i}=-\frac{1}{h}A_{k\mu}F^\mu_{\alpha i}-\frac{1}{2h}A_{\alpha j}\epsilon_{kij},\\
&H^i_{jk}=-\frac{1}{2}\epsilon_{ijk}.
\end{align}
We see that the connection in the fuzzy sphere is the QLC for that, while $\Gamma$ is the Christoffel symbol for the effective metric $\tilde g$ plus correction terms. We also see that $F$ is essentially the curvature of $A_{\mu i}$ if we think of this as an $su_2$ gauge field. 

\section{Real scalar field on the product as multiplets on spacetime} \label{seclap}
Here, we only consider the constant round fuzzy sphere case (\ref{round}). We calculate the Laplacian on $f$ as
\begin{align*}
\square f&=(\ ,\ )\nabla\extd f=(\ ,\ )\nabla(\del_i f s^i+\del_\alpha f\extd x^\alpha)\\
&=(\ ,\ )(\del_i f\nabla s^i+ (\extd \del_i f)\tens s^i+\del_\alpha f\nabla \extd x^\alpha+ (\extd \del_\alpha f)\tens \extd x^\alpha)\\
&=(\ ,\ )(\del_i f\nabla s^i+ (\del_j \del_i f s^j+\del_\alpha\del_i f\extd x^\alpha)\tens s^i+\del_\alpha f\nabla \extd x^\alpha+ (\del_i \del_\alpha f s^i+\del_\beta\del_\alpha f \extd x^\beta)\tens \extd x^\alpha)\\
&=\del_i f(\ ,\ )\nabla s^i+\del_i \del_j f\tilde h^{ij}+\del_\alpha\del_i f\tilde A^{\alpha i}+\del_\alpha f(\ ,\ )\nabla \extd x^\alpha+\del_i \del_\alpha f\tilde A^{i\alpha}+\del_\alpha\del_\beta f \tilde g^{\alpha\beta}.
\end{align*}
Substituting $\nabla\extd x^\alpha,\nabla s^i$, this becomes
\begin{align*}
\square f&=\del_\alpha\del_\beta f \tilde g^{\alpha\beta}+\del_i \del_j f\tilde h^{ij}+2\tilde A^{\alpha i}\del_\alpha\del_i f\\
&+\del_i f(E^i_{\alpha\beta}\tilde g^{\alpha\beta}+2B^i_{\alpha j}\tilde A^{\alpha j}+H^i_{jk}\tilde h^{jk})+\del_\mu f(-\Gamma^\mu_{\alpha\beta}\tilde g^{\alpha\beta}+2F^\mu_{\alpha j}\tilde A^{\alpha j}+D^\mu_{jk}\tilde h^{jk})
\end{align*}
which is the same result  as in \cite{ArgMa4}. By using $B,D,E,F,H$ from the end of Section~\ref{secQLC} and (\ref{Atilde}), (\ref{hgequation}), this can be written as
\begin{align*}
\square f&=  \tilde{g}^{\alpha \beta} \tilde\nabla_\alpha \partial_\beta f+\tilde{h}^{i j} \partial_j \partial_i f-\frac{1}{h}\tilde{g}^{\alpha \beta}\left( \nabla_\alpha A_{\beta i}+\frac{1}{h}\tilde g^{\mu \nu}A_{\beta k}A_{i\mu}\nabla_{[\alpha}A_{\nu]k}\right) \partial_i f-\frac{2}{h}\tilde g^{\alpha\beta}A_{\beta k} \partial_\alpha \partial_k f\\
&=  \tilde{g}^{\alpha \beta} \tilde\nabla_\alpha \partial_\beta f+\tilde{h}^{i j} \partial_j \partial_i f-\frac{1}{h}\tilde{g}^{\alpha \beta}\tilde \nabla_\alpha A_{\beta i}\partial_i f-\frac{2}{h}\tilde g^{\alpha\beta}A_{\beta k} \partial_\alpha \partial_k f\\
&=\tilde{g}^{\alpha \beta} \tilde\nabla_\alpha \partial_\beta f+\frac{1}{h}\partial_i \partial_i f+\frac{1}{h^2}\tilde g^{\mu\nu}A_{i\nu}A_{j\mu}\partial_j \partial_i f -\frac{1}{h}\tilde{g}^{\alpha \beta}\tilde \nabla_\alpha A_{\beta i}\partial_i f-\frac{2}{h}\tilde g^{\alpha\beta}A_{\beta k} \partial_\alpha \partial_k f\\
&=\tilde g^{\alpha\beta}(\tilde\nabla_{\alpha}-\frac{1}{h} A_{i\alpha}\del_i)(\tilde\nabla_{\beta}-\frac{1}{h} A_{j\beta}\del_j) f+\frac{1}{h}\partial_i \partial_i f\\
&=\tilde\square_A f +\frac{1}{h}\partial_i \partial_i f
\end{align*}
where 
\[ \tilde\square_A=\tilde g^{\alpha\beta}\tilde\nabla_{\alpha A}\tilde\nabla_{\beta A},\quad  \tilde\nabla_{\alpha A}=\tilde\nabla_{\alpha}-\frac{1}{h} A_{i\alpha}\del_i.\]

Therefore, denoting the different components in the orbital $l$ expansion (\ref{fuzzydec}) of $f$ now as fields $\phi_l$ on spacetime, and using (\ref{squareAl}), we obtain
\begin{align*}
\square f=\sum_{l=0}^\infty\left(\left(\tilde\square_{lA}-\frac{l(l+1)}{h}\right) \phi_l\right)_{i_1...i_l} y^{i_1}y^{i_2}...y^{i_l}, 
\end{align*}
where 
\begin{align*}
\tilde\square_{lA}=\tilde g^{\alpha\beta}\tilde\nabla^l_{\alpha A}\tilde\nabla^l_{\beta A},\quad \tilde\nabla^l_{\alpha A}=\tilde\nabla_{\alpha}+\frac{\imath}{h} A_{i\alpha}T^i_l
\end{align*}
and we understand $\tilde\square_{0A}:=\tilde\square=\tilde g^{\alpha\beta}\tilde\nabla_{\alpha}\tilde\nabla_{\beta}$ on $\phi_0$ where there are no spin indices. 

For the action , we have
\begin{align*}
S_f&=\int_M \extd^n x\sqrt{-|\tilde g|}\int_{\C_\lambda[S^2]} f\square f\\
&=\int_M\extd^n x\sqrt{-|\tilde g|}\int_{\C_\lambda[S^2]}\left(\sum_{l'=0}^\infty(\phi_{l'})_{j_1...j_{l'}}y^{j_1}y^{j_2}...y^{j_{l'}}\right)\left(\sum_{l=0}^\infty\left(\left(\tilde\square_{lA}-\frac{l(l+1)}{h}\right) \phi_l\right)_{i_1...i_{l}} y^{i_1}y^{i_2}...y^{i_{l}}\right)\\
&=\int_M\extd^n x\sqrt{-|\tilde g|}\sum_{l,l'=0}^\infty(\phi_{l'})_{j_1...j_{l'}}\left(\left(\tilde\square_{lA}-\frac{l(l+1)}{h}\right) \phi_{l}\right)_{i_1...i_{l}} \int_{\C_\lambda[S^2]} y^{j_1}y^{j_2}...y^{j_{l'}} y^{i_1}y^{i_2}...y^{i_{l}}\\
&=\int_M\extd^n x\sqrt{-|\tilde g|}\left( \sum_{l=0}^\infty(\phi_{l})_{j_1...j_{l}}\left(\left(\tilde\square_{lA}-\frac{l(l+1)}{h}\right) \phi_{l}\right)_{i_1...i_{l}}\alpha_l\delta^{j_1 {i}_{1}} \delta^{j_2 {i}_{2}}... \delta^{j_l {i}_{l}}\right)\\
&=\int_M\extd^n x\sqrt{-|\tilde g|} \sum_{l=0}^\infty \alpha_l\phi_l^T\left(\tilde\square_{lA}-\frac{l(l+1)}{h}\right) \phi_{l},
\end{align*}
where in the third expression,  the integral of $y^i$ consists of $\delta^{j_r i_s},\delta^{j_r j_s},\delta^{i_r i_s},\epsilon$, however only terms with $\delta^{j_r i_s}$ are left since the terms with $\epsilon$ vanish due to the totally symmetric coefficients in front of the integral and $\delta^{j_r j_s},\delta^{i_r i_s}$ vanish due to the traceless coefficients. Hence, given the symmetric coefficients, we have the fourth equality for some numerical constants $\alpha_l$ which could be computed. Here, $\alpha_0=1$.  

For example, we can see how this works on a piece of the initial two sums where $l,l'=1,2$. Then the relevant piece of the third expression is 
\begin{align*}
(\phi_{1}&)_{j_1}\left(\left(\tilde\square_{1A}-\frac{2}{h}\right) \phi_{1}\right)_{i_1}\int y^{j_1} y^{i_1}+(\phi_{1})_{j_1}\left(\left(\tilde\square_{2A}-\frac{2}{h}\right) \phi_{2}\right)_{i_1i_2}\int y^{j_1} y^{i_1}y^{i_2}\\
+&(\phi_{2})_{j_1j_2}\left(\left(\tilde\square_{1A}-\frac{2}{h}\right) \phi_{1}\right)_{i_1}\int y^{j_1} y^{j_2}y^{i_1}+(\phi_{2})_{j_1j_2}\left(\left(\tilde\square_{2A}-\frac{2}{h}\right) \phi_{2}\right)_{i_1i_2}\int y^{j_1} y^{j_2}y^{i_1}y^{i_2}.
\end{align*}
From Section~\ref{secfuzzy}, the first term is
\begin{align*}
(\phi_{1})_{j_1}\left(\left(\tilde\square_{1A}-\frac{2}{h}\right) \phi_{1}\right)_{i_1}\frac{1-\lambda^2}{3}\delta^{j_1 i_1}=(\phi_{1})_{i_1}\left(\left(\tilde\square_{1A}-\frac{2}{h}\right) \phi_{1}\right)_{i_1}\frac{1-\lambda^2}{3}
\end{align*}
so that
\[\alpha_1=\frac{1-\lambda^2}{3}.\]
The second term is
\begin{align*}
(\phi_{1})_{j_1}\left(\left(\tilde\square_{2A}-\frac{2}{h}\right) \phi_{2}\right)_{i_1i_2}{\imath\lambda(1-\lambda^2)\over 3}\eps^{j_1 i_1 i_2}=0.
\end{align*}
Similarly, the third term is $0$, and the fourth term is
\begin{align*}
(\phi_{2}&)_{j_1j_2}\left(\left(\tilde\square_{2A}-\frac{2}{h}\right) \phi_{2}\right)_{i_1i_2} {1-\lambda^2\over 15}\left((1+\lambda^2)(\delta^{j_1j_2}\delta^{i_1i_2}+\delta^{j_1i_2}\delta^{j_2i_1})+(1-9\lambda^2)\delta^{j_1i_1}\delta^{j_2i_2} \right)\\
&={1-\lambda^2\over 15}\left((1+\lambda^2)(\phi_{2})_{i_2i_1}\left(\left(\tilde\square_{2A}-\frac{2}{h}\right) \phi_{2}\right)_{i_1i_2}+(1-9\lambda^2)(\phi_{2})_{i_1i_2}\left(\left(\tilde\square_{2A}-\frac{2}{h}\right) \phi_{2}\right)_{i_1i_2}\right)\\
&={2(1-\lambda^2)(1-4\lambda^2)\over 15}(\phi_{2})_{i_1i_2}\left(\left(\tilde\square_{2A}-\frac{2}{h}\right) \phi_{2}\right)_{i_1i_2}
\end{align*}
from which we can read off
\[\alpha_2={2(1-\lambda^2)(1-4\lambda^2)\over 15}.\]

Finally, we rescale the component fields $\phi_l$ to absorb these constants, so that the real scalar field action on the product can be written as
\begin{equation}\label{Sf}S=\int_M\extd^n x\sqrt{-|\tilde g|}\sum_{l=0}^\infty \phi_l^T\left(\tilde\square_{lA}-\frac{l(l+1)}{h}\right) \phi_{l}.\end{equation}

\section{Product Ricci scalar and the emergence of the YM action}\label{secR}
In this section we calculate Ricci scalar for the product QRG, again in the case (\ref{round}) of a constant round fuzzy sphere. We first calculate the classical Ricci scalar with respect to $\tilde g_{\mu\nu}$ (but in QRG conventions where we use $-1/2$ of the usual value),
\begin{align*}
\tilde R_M&=\frac{1}{2}\tilde{g}^{\alpha \beta}(-\partial_\mu \tilde\Gamma_{\alpha \beta}^\mu+\partial_\alpha \tilde\Gamma_{\mu \beta}^\mu+\tilde\Gamma_{\alpha \nu}^\mu \tilde\Gamma_{\mu \beta}^\nu-\tilde\Gamma_{\mu \nu}^\mu \tilde\Gamma_{\alpha \beta}^\nu)\\
&=\frac{1}{2}\tilde{g}^{\alpha \beta}\left(-\partial_\mu \Gamma_{\alpha \beta}^\mu+\partial_\alpha \Gamma_{\mu \beta}^\mu+\Gamma_{\alpha \nu}^\mu \Gamma_{\mu \beta}^\nu-\Gamma_{\mu \nu}^\mu \Gamma_{\alpha \beta}^\nu+\frac{1}{h} \nabla_\alpha\left(A_{\mu i} F_{\beta i}^\mu\right)-\frac{2}{h} A_{\alpha i} \nabla_\mu F_{\beta i}^\mu\right. \\
& \left.-\frac{2}{h} (\nabla_\mu A_{\alpha i}) F_{\beta i}^\mu+\frac{1}{h^2}\left(A_{\mu j} A_{\nu i} F_{\alpha i}^\mu F_{\beta j}^\nu+A_{\alpha i} F_{\nu i}^\mu A_{\beta j} F_{\mu j}^\nu+2(F^\mu_{j\beta}F^\nu_{i\mu}-F^\mu_{i\beta}F^\nu_{j\mu})A_{\nu j}A_{\alpha i}\right)\right)
\end{align*}
Then we use (\ref{RR}) to compute  the Ricci scalar on the product as
\begin{align*}
&R=-\frac{1}{4}\tilde h^{ii}+\frac{1}{2}\tilde g^{\alpha\beta}\left(-\partial_\mu \Gamma_{\alpha \beta}^\mu+\partial_\alpha \Gamma_{\mu \beta}^\mu+\Gamma_{\alpha \nu}^\mu \Gamma_{\mu \beta}^\nu-\Gamma_{\mu \nu}^\mu \Gamma_{\alpha \beta}^\nu+\frac{1}{h}\nabla_\alpha(A_{i\mu}F^\mu_{i\beta})+\frac{1}{h^2}A_{j\mu}A_{i\nu}F^\mu_{i\alpha}F^\nu_{j\beta}\right.\\ &\left. -\frac{2}{h}A_{\alpha i}\nabla_\mu F^\mu_{\beta i}+\frac{2}{h^2}A_{i\beta}A_{\nu j}(F^\mu_{\alpha j}F^\nu_{\mu i}-F^\mu_{\alpha i}F^\nu_{\mu j})+\frac{1}{h^2}A_{i\mu}A_{k\alpha}F^\mu_{\beta j}\epsilon_{ijk}-\frac{1}{h}\nabla_{(\alpha}A_{\mu)i}F^\mu_{\beta i}+\frac{1}{2h^2}A_{\alpha i}A_{\beta i}\right)\\
&+\frac{1}{2}\tilde h^{ij}F^\mu_{\nu i}F^\nu_{\mu j}.
\end{align*}
Substituting $\tilde R_M$,  $R$ can be written as
\begin{align*}
R&=\tilde R_M+\frac{1}{2h}\tilde g^{\alpha\beta}\left(\del_{[\mu} A_{\alpha] i} F_{\beta i}^\mu-\frac{1}{h}A_{\alpha i}A_{\beta j}F^\nu_{\mu j}F^\mu_{\nu i}-\frac{1}{h}A_{j\mu}A_{k\alpha}F^\mu_{\beta i}\epsilon_{ijk}\right)\\
&\quad +\frac{1}{2}\tilde h^{ij}F^\mu_{\nu i}F^\nu_{\mu j}+\frac{1}{4h^2}\tilde g^{\alpha\beta}A_{\alpha i}A_{\beta i}-\frac{1}{4}\tilde h^{ii}\\
&=\tilde R_M+(\frac{1}{2}\tilde h^{ij}-\frac{\delta^{ij}}{h}-\frac{1}{2h^2}\tilde g^{\alpha\beta}A_{\alpha i}A_{\beta j})F^\mu_{\nu i}F^\nu_{\mu j}+\frac{1}{4h^2}\tilde g^{\alpha\beta}A_{\alpha i}A_{\beta i}-\frac{1}{4}\tilde h^{ii}
\end{align*}
Finally, using (\ref{hgequation}), $R$ becomes
\begin{align*}
R=&\tilde R_M-\frac{3}{4h}-\frac{1}{2h}F^\mu_{\nu i}F^\nu_{\mu i}.
\end{align*}

Next, we identify the curvature of $A$ as
\begin{align}\label{Fgauge}
F^i_{\alpha\beta}=2\tilde g_{\alpha \mu}F^\mu_{\beta i}=\del_{[\alpha}A_{\beta]i}-\frac{1}{h}A_{\alpha j}A_{\beta k}\epsilon_{ijk},\quad F^{i\alpha\beta}=\tilde g^{\alpha\mu}\tilde g^{\beta\nu}F^i_{\mu\nu}
\end{align}
so that
\begin{align}\label{Ricci}
R=\tilde R_M-\frac{3}{4h}+\frac{1}{8h}F^i_{\alpha\beta}F^{i\alpha\beta}.
\end{align}
We see that the cross terms produces a Yang-Mills action. The second term is the constant Ricci scalar of the scaled fuzzy sphere. 

\section{General $h_{ij}$ and Liouville-sigma model action}\label{secgen}

We have so far focussed on the fibre being a constant round fuzzy sphere. However, the metric on the fuzzy sphere does not have to be the round metric and could also vary over spacetime. This now appears as an additional field $h_{ij}$ on spacetime and in this section we give the much more complicated analysis of the scalar Laplacian and Ricci scalar on the product in terms of how these appear on spacetime. In preparation for our results, we identify now the more relevant gauge field and its curvature as
\begin{equation}\label{tildeA}\tilde{A}_{\alpha i}:=h^{i j}A_{\alpha j},\quad \tilde F^i_{\alpha\beta}=\del_{[\alpha}\tilde A_{\beta]i}-\tilde A_{\alpha j}\tilde A_{\beta k}\epsilon_{ijk}, \quad \tilde F^{i\alpha\beta}=\tilde g^{\alpha\mu}\tilde g^{\beta \nu}\tilde F^i_{\mu\nu}.\end{equation}
The associated covariant derivative when acting on tensors on spacetime which are scalars on the fuzzy sphere is
\begin{equation}\label{covAh} \tilde\nabla_{A\alpha }:=\tilde\nabla_{\alpha}-\tilde A_{i\alpha}\del_i\end{equation}
where $\tilde\nabla_\alpha$ is the Levi-Civita connection  for the physical metric $\tilde g_{\mu\nu}$. If this acts on a spacetime scalar then we would just use $\del_\alpha$. The same covariant derivative when acting on $h_{ij}$, as this has indices that rotate under the $SU(2)$ symmetry of the fuzzy sphere but is constant on the fuzzy sphere, is 
\begin{equation}\label{nablah} \tilde\nabla_{A \alpha} h_{ij}:=\del_\alpha h_{ij}+\tilde A_{\alpha k}h_{l(i}\epsilon_{j)kl}.\end{equation}
We won't need to act on field with both spacetime and fuzzy sphere tensor indices but if we did we would use such expressions with $\tilde\nabla_\alpha$ in place of $\del_\alpha$ here. We also won't need to act on fields which have fuzzy sphere indices and vary on the fuzzy sphere, but if we did we would also have another term with $\del_i$ acting.  Recall that $h_{ij}$ is forced by the QRG to be constant on the fuzzy sphere.

 We now consider the general product metric (\ref{prodmetric}). We already solved for the QLC for the general case, with coefficients $E, F, B, D, H,\Gamma$ given in (\ref{sE}), (\ref{F2}), (\ref{B2}), (\ref{sD}), (\ref{sH}), (\ref{G2}) respectively. 
 
 \begin{proposition}\label{proplaph} The QRG Laplacian $\square=(\ ,\ )\nabla\extd$ acting on a scalar field $f$ on the product is 
  \[ \square f=\tilde\square_A f+ \square_h f+\frac{1}{2}\tilde g^{\alpha\beta}h^{ij}(\del_\alpha h_{ij})\tilde\nabla_{A\beta}f,\]
where   $\tilde\square_A=\tilde g^{\alpha\beta}\tilde\nabla_{A\alpha}\tilde\nabla_{A\beta}$ and  $\square_h=h^{ij}\del_i\del_j$ is the Laplacian on the fuzzy sphere.\end{proposition}
\proof Putting in the expressions for $E, F, B, D, H,\Gamma$, we eventually obtain
\begin{align*}
\square f&=\tilde g^{\alpha\beta}\tilde\nabla_\alpha\del_\beta f+h^{ij}\del_i\del_j f+h^{ik}h^{lj}\tilde g^{\alpha\beta}A_{k\alpha}A_{l\beta}\del_i\del_j f-2\tilde{g}^{\alpha \beta}h^{i j}A_{j \beta}\del_\alpha\del_i f\\
&\quad -\del_i f\tilde g^{\alpha\beta}\tilde\nabla_{\alpha} (A_{ \beta j}h^{ij})+\frac{1}{2}(\del_\beta f- A_{l\beta}h^{kl}\del_k f)\tilde g^{\alpha\beta}h^{ij}\del_\alpha h_{ij}\\
&=\tilde g^{\alpha\beta}(\tilde\nabla_{\alpha}-h^{ik} A_{i\alpha}\del_k)(\tilde\nabla_{\beta}-h^{lj} A_{l\beta}\del_j) f+h^{ij}\partial_i \partial_j f+\frac{1}{2}(\del_\beta f- A_{l\beta}h^{kl}\del_k f)\tilde g^{\alpha\beta}h^{ij}\del_\alpha h_{ij}
\end{align*}
which we write as stated. Here $\tilde\nabla_\beta =\del_\beta$ when acting on $f$, as this is a scalar on spacetime. We also recognise the Laplacian on the fuzzy sphere given in  (\ref{lapfuzzy}).\endproof

We see the emergence of an interaction coupling between the log-derivative of $h_{ij}$ and  $\tilde\nabla f$. This suggests to work with $\Phi=\ln (h)$ as a matrix, which makes sense if $h_{ij}$ is positive as a real symmetric matrix, which we now assume for the fuzzy sphere to have a spatial signature in our conventions. We then define
 \[\Phi^i_{\alpha j}:=h^{ik}\tilde\nabla_{A\alpha} h_{kj},\quad {\rm Tr}(\Phi_\alpha)= h^{ij}\del_\alpha h_{ij}=\del_\alpha {\rm Tr}(\Phi)=\del_\alpha\ln(\det(h))\]
so that the result of Proposition~\ref{proplaph} can be  written compactly as
\begin{equation}\label{lapgen} \square=\tilde\square_A+ \square_h +\frac{1}{2}\tilde g^{\alpha\beta}{\rm Tr}(\Phi_\alpha)\tilde\nabla_{\beta A} f=\tilde\square_A+ \square_h +\frac{3}{2}\tilde g^{\alpha\beta}(\del_\alpha \varphi)\tilde\nabla_{\beta A} f,\end{equation}
where $\varphi:=\ln(\det(h))/3$. We see a log-derivative coupling to the fuzzy sphere metric volume as it varies on spacetime. This is broadly similar to such log coupling terms in some of the models in \cite{ArgMa4,LiuMa2}. We now look at the Ricci scalar of the product QRG.

\begin{proposition}\label{propRprodh} For general $h_{ij}$, the Ricci scalar on the product is 
\begin{align*}
\ R=&\tilde R_M+R_h+\frac{1}{8}h_{ij}\tilde F^i_{\mu\nu}\tilde F^{j\mu\nu}+\frac{1}{2}\tilde{g}^{\alpha \beta}\tilde\nabla_\alpha {\rm Tr}(\Phi_\beta)+\frac{1}{8}\tilde{g}^{\alpha \beta}\big({\rm Tr}(\Phi_\alpha\Phi_\beta)+{\rm Tr}(\Phi_\alpha){\rm Tr}(\Phi_\beta)\big)
\end{align*}
where $\tilde R_M$ is the  Ricci scalar on $M$ for the physical metric $\tilde g$ and
\[R_{h}={e^{-\Tr(\Phi)}\over 2}\Big(\Tr(e^{2\Phi})-{1\over 2}\Tr(e^\Phi)^2\Big) \]
is the Ricci scalar on the fuzzy sphere regarded as a potential term for $\Phi$.
\end{proposition}
\proof
We  use (\ref{RR}) to calculate the general Ricci scalar on the product, which gives
\begin{align*}
R=&\frac{1}{2}\Big(-\nabla_\alpha B_{\beta k}^k-F_{\mu k}^\mu E_{\alpha \beta}^k+2F_{\alpha k}^\mu E_{\mu \beta}^k+B_{\alpha k}^n B_{\beta n}^k-E_{\alpha \beta}^n H_{k n}^k-\Gamma_{\alpha \beta}^\nu \Gamma_{\mu \nu}^\mu \\
& +\Gamma_{\mu \beta}^\nu \Gamma_{\nu \alpha}^\mu-\Gamma_{\alpha \beta, \mu}^\mu+\Gamma_{\mu \beta, \alpha}^\mu-A_{j \beta} h^{i j}\big(2 \nabla_\mu F_{\alpha i}^\mu-\nabla_\alpha F_{\mu i}^\mu+2D_{i k}^\mu E_{\mu \alpha}^k-2 F_{\mu k}^\mu B_{\alpha i}^k \\
& + 2 F_{\alpha k}^\mu B_{\mu i}^k-2 F_{\alpha i}^\mu B_{\mu k}^k+2B_{\alpha k}^n H_{ni}^k-2 B_{\alpha i}^n H_{k n}^k-H_{k i, \alpha}^k\big)\Big) \tilde{g}^{\alpha \beta} \\
& +\frac{1}{2}\left(\nabla_\mu D_{i j}^\mu+F_{\mu i}^\nu F_{\nu j}^\mu+2D_{i k}^\mu B_{\mu j}^k-D_{i j}^\mu B_{\mu k}^k-F_{\mu k}^\mu H_{i j}^k+H_{ni}^k H_{k j}^n-H_{i j}^n H_{k n}^k\right) \tilde{h}^{i j}.
\end{align*}
Next, in order to substitute $E, F, B, D, H,\Gamma$ more easily, we note from their expression in Section~\ref{secQLC} that \begin{align*}B^k_{\mu j}&=-h^{rk}A_{r\nu}F^\nu_{\mu j}+\frac{1}{2}h^{rk}(A_{l\mu}\epsilon_{jrl}-\del_\mu h_{jr})\\
H^k_{ni}&=\frac{1}{2}h^{rk}h_{l(i}\epsilon_{n)rl}-h^{rk}A_{r \mu} D_{n i}^\mu-\frac{1}{2}\epsilon_{kni}\\
\tilde\Gamma^\mu_{\alpha\beta}&=\Gamma^\mu_{\alpha\beta}+h^{ij}F^\mu_{j(\alpha}A_{\beta) i}-h^{pi}h^{qj}A_{\alpha p}A_{\beta q}D^{\mu}_{ij}\\
\tilde\Gamma^\mu_{\mu\beta}&=\Gamma^\mu_{\mu\beta}-B^{l}_{\beta l}-\frac{1}{2}h^{ij}\del_\beta h_{ij}\\
F_{\mu i}^\mu&=-H_{ki}^k\\
F^\mu_{\alpha i}&=-\frac{1}{2}h_{ij}\tilde g^{\mu\nu}\tilde F^j_{\alpha\nu}+\tilde A_{k\alpha}D^\mu_{ik}.\end{align*}
Using these, we first calculate $\tilde R_M$ as
\begin{align*}
\tilde R_M=&\frac{1}{2}\tilde{g}^{\alpha \beta}(-\partial_\mu \tilde\Gamma_{\alpha \beta}^\mu+\partial_\alpha \tilde\Gamma_{\mu \beta}^\mu+\tilde\Gamma_{\alpha \nu}^\mu \tilde\Gamma_{\mu \beta}^\nu-\tilde\Gamma_{\mu \nu}^\mu \tilde\Gamma_{\alpha \beta}^\nu)\\
=&\frac{1}{2}\tilde{g}^{\alpha \beta}(-\partial_\mu \Gamma_{\alpha \beta}^\mu+\partial_\alpha \Gamma_{\mu \beta}^\mu+\Gamma_{\alpha \nu}^\mu \Gamma_{\mu \beta}^\nu-\Gamma_{\mu \nu}^\mu \Gamma_{\alpha \beta}^\nu-\nabla_\alpha B^{l}_{\beta l}+2h^{ij}A_{j\beta}(-\nabla_\mu F^\mu_{i\alpha}+F^\nu_{i\alpha}B^{l}_{\nu l})-\frac{1}{2}\tilde\nabla_\alpha\text{Tr}(\Phi_\beta)\\
&-2F^\mu_{j\alpha}\nabla_\mu (h^{ij}A_{\beta i})+h^{pi}h^{qj}A_{\alpha p}A_{\beta q}(\nabla_\mu D^{\mu}_{ij}+F^\mu_{i\nu}F^\nu_{j\mu}-B^{l}_{\nu l}D^{\nu}_{ij})+D^{\mu}_{ij}\nabla_\mu(h^{pi}h^{qj}A_{\alpha p}A_{\beta q})\\
&+h^{mn}h^{ij}A_{\mu n}A_{\nu i}F^\mu_{j\alpha}F^\nu_{m\beta}+2h^{nm}h^{ij}A_{\alpha i}A_{\mu n}F^\mu_{j\nu}F^\nu_{m\beta}\\
&-2h^{ij}h^{rm}h^{sn}A_{\alpha i}A_{\mu r}A_{\beta s}D^{\nu}_{mn}F^\mu_{j\nu}-2h^{ij}h^{rm}h^{sn}A_{\mu r}A_{\nu i}A_{\beta s}D^{\nu}_{mn}F^\mu_{j\alpha}\\
&+h^{pi}h^{qj}h^{rm}h^{sn}A_{\mu r}A_{\beta s}A_{\alpha p}A_{\nu q}D^{\mu}_{ij}D^{\nu}_{mn}). 
\end{align*}
Comparing terms with $R$ and after a great deal of algebra, we eventually find  
\begin{align*}
R=&\tilde R_M+R_h+\frac{1}{8}h_{ij}\tilde F^i_{\mu\nu}\tilde F^{j\mu\nu}+\frac{1}{2}\tilde{g}^{\alpha \beta}\tilde\nabla_\alpha (h^{ij}\del_\beta h_{ij})-\frac{1}{8}\tilde{g}^{\alpha \beta}\del_\alpha h^{ij}\del_\beta h_{ij}+\frac{1}{8}\tilde{g}^{\alpha \beta}(h^{ij}\del_\alpha h_{ij})(h^{kl}\del_\beta h_{kl})\\
&+\frac{1}{2}\tilde{g}^{\alpha \beta}\tilde A_{\alpha l}h_{ni}\epsilon_{jln}\del_\beta h^{ij}-\frac{1}{2}\tilde{g}^{\alpha \beta}\tilde A_{\alpha i}\tilde A_{\beta i}-\frac{1}{4}\tilde{g}^{\alpha \beta}h^{ml}h_{n k}\tilde A_{\alpha i}\tilde A_{\beta j}\epsilon_{k l i}\epsilon_{jmn},
\end{align*}
where we recognise 
\begin{equation}\label{Rh}R_{h}=-\frac{1}{2}(h^{i i}+\frac{1}{4} h^{i j}h^{lk}h_{mn}\epsilon_{imk}\epsilon_{jln})\end{equation}
as an equivalent form of the Ricci scalar on the fuzzy sphere in (\ref{Rfuzzy}). We also recognised the Yang-Mills action with the metric $h_{ij}$ for contraction of the Lie algebra indices.

Finally, we recall the covariant derivative (\ref{nablah}) and, noting that $h^{ij}\tilde\nabla_{A\mu}h_{ij}=h^{ij}\del_\mu h_{ij}$ as explained above and 
\[\del_\alpha h^{ij}=-h^{il}h^{jk}\del_\alpha h_{lk}=-h^{il}h^{jk}(\tilde\nabla_{A\alpha} h_{lk}-\tilde A_{\alpha p}h_{q(l}\epsilon_{k)pq}),\]
we can write our result as
\begin{align}\label{Rprodh}\nonumber R=&\tilde R_M+R_h+\frac{1}{8}h_{ij}\tilde F^i_{\mu\nu}\tilde F^{j\mu\nu}+\frac{1}{2}\tilde{g}^{\alpha \beta}\tilde\nabla_\alpha (h^{ij}\tilde\nabla_{A\beta} h_{ij})\\
&\quad+\frac{1}{8}\tilde{g}^{\alpha \beta}h^{ik}h^{jl}\tilde\nabla_{A\alpha} h_{kl}\tilde\nabla_{A\beta} h_{ij}+\frac{1}{8}\tilde{g}^{\alpha \beta}(h^{ij}\tilde\nabla_{A\alpha} h_{ij})(h^{kl}\tilde\nabla_{A\beta} h_{kl}).
\end{align}
We then write this in terms of $\Phi$ as stated.
\endproof

The fourth term in the result is a total derivative and does not contribute to the Einstein-Hilbert action on the product (this only needs an integral over $M$ since none of the fields in $R$ vary over the fuzzy sphere). We see a  non-linear sigma model kinetic term $\tilde g^{\alpha\beta}{\rm Tr}(\Phi_\alpha\Phi_\beta)$ but the model is rather different in character from well-known models such as the Wess-Zumino model on a group, since $h_{ij}$ is not a group valued field but a symmetric matrix valued field, and the topological term (which does not apply in our context) is replaced by other quadratic terms and the fuzzy sphere curvature $R_h$. We see that the latter contributes a matrix version of a Liouville-like potential.  In fact,  the final result (\ref{Rprodh}) is remarkably of the same form as obtained from a classical Kaluza-Klein ansatz in  \cite[Chap.~4]{Coq} for classical $S^3$ fibre equipped with a right-invariant metric, although this not our case and also we are not making an ansatz but merely analysed the full content of QRG on the product.

\subsection{Restriction to the non-constant round metric case}

A special case, which is intermediate between the most general case above and the constant round metric case in earlier sections, is $h_{ij}=h\delta_{ij}$ for $h$ now a positive-valued scalar field on spacetime. Here $h$ corresponds to the area of the fuzzy sphere as we move around spacetime. The above scalar field $\varphi$ defined as $\ln(\det(h))/3$ of the metric matrix now just reduces to 
\[ \varphi=\ln(h)\]
 as the corresponding scalar Liouville field. Then the Laplacian in Proposition~\ref{proplaph} simplifies to
\begin{equation}\label{laphvar}\square f=(\tilde\square_A + e^{-\varphi}\partial_i \partial_i) f+\frac{3}{2}\tilde g^{\alpha\beta}(\del_\alpha \varphi)\tilde\nabla_{\beta A}f,\end{equation}
which generalises the formulae at the end of Section~\ref{secQLC} to allow for non-constant $h$. The only difference is again an extra coupling between the log derivative of $h$ and the covariant derivative of $f$ (where the fuzzy sphere dependence of $f$, decomposed under orbital spin, now appears on spacetime as fields of different internal spin).

Similarly, the Ricci scalar on the product in Proposition~\ref{propRprodh} simplifies up to a total derivative to 
\begin{equation}\label{Rvarphi} R=\tilde R_M +\frac{e^\varphi}{8}\tilde F^i_{\alpha\beta}\tilde F^{i\alpha\beta}+ {3\over 2}\left(\tilde g^{\alpha\beta}(\del_\alpha\varphi)(\del_\beta\varphi) -{1\over 2} e^{-\varphi}\right).\end{equation}
 This analysis also holds in the classical limit and we see the emergence of a Liouville action for $\varphi$ as expected. Scalar fields on the product couple to the derivative of $\varphi$ as in (\ref{laphvar}) and we also have $e^{-\varphi}$  scaling the $l(l+1)$ mass terms in the multiplet expansion there and $e^{\varphi}$ scaling the geometric Yang-Mills action in (\ref{Rvarphi}).  Because of the latter, the physical matching with usual Yang-Mills needs $\varphi$ constant as could presumably happen in its vacuum configuration.

As an independent check  on all our calculations, it is instructive to also obtain the result in this case directly. Thus, the various quantities that make up the QLC on the product in this context reduce to
\begin{align*}
&\tilde\Gamma^\sigma_{\mu\nu}=\Gamma^\sigma_{\mu\nu}+\frac{1}{h}A_{i(\mu}F^\sigma_{\nu) i}-\frac{1}{2h^2}\tilde g^{\rho\sigma}A_{\mu i}A_{\nu i}\del_\rho h,\\
&E_{\alpha \beta}^i=-\frac{1}{2h}\nabla_{(\alpha} A_{ \beta)i},\\
&B^k_{\mu i}=-\frac{1}{h}A_{k\nu}F^\nu_{\mu i}+\frac{1}{2h}A_{l\mu}\epsilon_{ikl}-\frac{1}{2h}\delta^k_{i}\del_\mu h,\\
&H^k_{ni}=-\frac{1}{2h}A_{k\mu}\tilde g^{\mu\alpha}\delta_{in}\del_\alpha h-\frac{1}{2}\epsilon_{kni},\\
&D^\mu_{ij}=\frac{1}{2}\tilde g^{\mu\alpha}\delta_{ij}\del_\alpha h,\\
&F^\mu_{\alpha i}=\frac{1}{2}\tilde g^{\beta\mu}(-\del_{[\alpha}A_{\beta]i}-\frac{1}{h}\epsilon_{ijk}A_{j\beta}A_{\alpha k})+\frac{1}{2}\tilde g^{\beta\mu}A_{i\beta}\del_\alpha \ln(h)
\end{align*}
which after a lengthy computation from the general expression given at the start of the proof of Proposition~\ref{propRprodh} gives
\begin{align*}
R&=\tilde R_M-\frac{3}{4h}+\frac{1}{8h}F^i_{\alpha\beta}F^{i\alpha\beta}+\frac{3}{2h}\tilde g^{\alpha\beta}\tilde \nabla_\alpha \del_\beta h+
\frac{1}{2h^2}\tilde g^{\alpha\beta}\tilde g^{\mu\nu} A_{\beta i}(\del_\mu h)(F^i_{\alpha\nu}+\frac{1}{2h}A_{i[\alpha}\del_{\nu]} h),\end{align*}
where we identify the Yang-Mills curvature of $A$ as 
\[ F^i_{\alpha\beta}=\del_{[\alpha}A_{\beta]i}-\frac{1}{h}A_{\alpha j}A_{\beta k}\epsilon_{ijk},\quad F^{i\alpha\beta}=\tilde g^{\alpha\mu}\tilde g^{\beta\nu}F^i_{\mu\nu}\]
now coming from just the first part of the expression for $F^\mu_{\beta i}$, i.e. without the $\del_\beta\ln(h)$ term. Finally, the geometric gauge field  in our context and its curvature are
\[ \tilde A_{\mu i} :=  {A_{\mu i}\over h};\quad  \tilde F^i_{\alpha\beta}:=\del_{[\alpha}\tilde A_{\beta]i}-\tilde A_{\alpha j}\tilde A_{\beta k}\epsilon_{ijk}={1\over h}\left(F^i_{\alpha\beta}+{1\over h}A_{i[\alpha}\del_{\beta]}h\right).\]
We also define $\tilde F^{i\alpha\beta}$ by raising indices with $\tilde g$ as before. Then
\begin{equation}\label{Rvarh} R=\tilde R_M-\frac{3}{4h}+\frac{h}{8}\tilde F^i_{\alpha\beta}\tilde F^{i\alpha\beta}+\frac{3}{2h}\tilde g^{\alpha\beta}\tilde \nabla_\alpha \del_\beta h+\frac{3}{2h^2}\tilde g^{\alpha\beta}(\del_\alpha  h)( \del_\beta h)\end{equation}
in full agreement with (\ref{Rprodh}) specialised to $h_{ij}=h\delta_{ij}$. This then converts to (\ref{Rvarphi}) in terms of $\varphi=\ln(h)$. 

\section{Concluding remarks}\label{seccon}

Putting together the results of Sections~\ref{seclap},~\ref{secR}, we obtained for the constant round metric $h_{ij}=h\delta_{ij}$ on the fuzzy sphere that the total action $S=\int R+f\square f$  for gravity and a massless real scalar field on the product  appears on spacetime $M$ as
\begin{align*}
S&=\int_M\extd^n x\sqrt{-|\tilde g|}\left(\tilde R_M-\frac{3}{4h}+\frac{1}{8h}F^i_{\alpha\beta}F^{i\alpha\beta}+\sum_l \phi_l^T\left(\tilde\square_{lA}-\frac{l(l+1)}{h}\right) \phi_{l}\right)
\end{align*}
i.e. a tower of real scalar fields $\{\phi_l\}$ in the $su_2$ integer spin $l$ representation and with square mass $\frac{l(l+1)}{h}$ coupled with a massless Yang-Mills field $A_{\mu i}$, and gravity. Here $h$ is a constant but we also saw in Section~\ref{secgen} that if we allow the fuzzy sphere metric $h_{ij}$ to become dynamical then it appears as a Liouville-sigma model, or a scalar Liouville action if we restrict to metrics that are round but the scalar $h$ is no longer constant. 

Our result for the round metric and generic $\lambda$ is the same answer as the Kaluza-Klein ansatz for classical $S^2$ fibre (which {\em supposes} that the inverse metric has cross terms of a particular form given by a vector field on $M$ and Killing vectors on the fibre, in our case given by the  $\del^i$). But the big difference is that the product form of the metric (\ref{prodmetric}) is not an ansatz and instead forced by a very mild assumption that the fibre (quantum) geometry commutes with the geometry on $M$, and the tight constraints of standard QRG. This happened because invertibility in the standard form of QRG \cite{BegMa} forces $\cg$ to be central which, since the $s^i$ are a central basis and since $\C_\lambda[S^2]$, as well as the reduced $c_\lambda[S^2]$ for the discrete series of $\lambda$, all have trivial centre, then forces (\ref{prodmetric}) as noted in \cite{ArgMa4}. We saw how this extends to the QLC also. Hence, our analysis derives the Kaluza-Klein ansatz here as a consequence of the rigidity of noncommutative geometry not visible in the classical limit $\lambda=0$.  We also have the new possibility of truncation of modes for $\lambda$ in the discrete series,  and we have a more general real symmetric matrix field $h_{ij}$ if we do not specialise to the round metric on the fuzzy sphere. 

Proceeding in the constant round metric case, the matching to Yang-Mills with $SU(2)$ gauge coupling $g$ then goes the same as for classical Kaluza-Klein theory. We set $\hbar=c=1$, then the physically normalised   Einstein-Hilbert action from (\ref{Ricci}) is 
\[S=-\frac{1}{8\pi G}\int_M\extd^n x \sqrt{-|\tilde g|}\tilde R=-\frac{1}{8\pi G}\int_M\extd^n x \sqrt{-|\tilde g|}\left(\tilde R_M-\frac{3}{4h}+\frac{1}{8h}F^i_{\alpha\beta}F^{i\alpha\beta}\right)\]
on remembering  that our $\tilde R$ is $-1/2$. This then gives the correct Einstein-Hilbert action in $M$. We ignore the constant $-{3\over 4}$ from the curvature on the fuzzy sphere and then to get the usual $-{1\over 4}||\bar F||^2$ for the physically normalised field $\bar A^i_{\mu}$ with curvature $\bar F^i{}_{\mu\nu}$, we need 
\[ F^i_{\alpha\beta}=-\sqrt{16\pi G h}\bar F^i_{\alpha\beta}.\]
Then substituting (\ref{Fgauge}) and $\bar F^i_{\alpha\beta}=\del_{[\alpha}\bar A^i_{\beta]}+g \epsilon^{ijk}\bar A^j_{\alpha}\bar A^k_{\beta }$ into the above relationship, we have
\begin{align*}
\del_{[\alpha}A_{\beta]i}-\frac{1}{h}A_{\alpha j}A_{\beta k}\epsilon_{ijk}=-\sqrt{16\pi G h}\left(\del_{[\alpha}\bar A^i_{\beta]}+g \epsilon^{ijk}\bar A^j_{\alpha}\bar A^k_{\beta }\right).
\end{align*}
Matching the linear and quadratic terms on both sides, we obtain
\[A_{\alpha i}=-\sqrt{16\pi G h}\bar A^i_{\alpha},\quad g=\sqrt{16\pi G \over h}.\]
The value of $g$ here is also what we get upon matching $\tilde\nabla^l_{\alpha A}=\tilde\nabla_{\alpha}+\frac{\imath}{h} A_{\alpha i}T^i_l$ with the Yang-Mills covariant derivative $\nabla_{\alpha A}=\nabla_{\alpha}-\imath g \bar A^i_{\alpha}T^i$. 

Now if we suppose, for comparison purposes only, that $\bar A$ are weak $SU(2)$ gauge bosons in the Standard Model, then we would want
\[h=\frac{16\pi G}{g^2}=\frac{4G}{\alpha_W}=120G\approx(11\lambda_P)^2,\quad A_{\alpha i}=-\frac{16\pi G}{g}\bar A^i_\alpha,\]
where $\alpha_W=\frac{g^2}{4\pi}\approx\frac{1}{30}$ is the relevant version of the finite structure constant for the weak force. This is necessarily the same as for the classical Kaluza-Klein ansatz but the reader should note that  if the sphere at each point of spacetime is 11 Planck lengths in radius, it is very unreasonable to take here a classical sphere, as one does not expect continuum geometry at the Planck scale. It is more reasonable that it could be modelled, after including relevant quantum gravity corrections, as a highly fuzzy sphere far from $\lambda=0$. The other extreme in the discrete series case would be $\lambda =1/(2j+1)$ with $j=0,1/2, 1$ for example. The same complaint applies to the original Kaluza-Klein work where, to match electromagnetism, one needs $h=(23\lambda_P)^2$ so that the $S^1$ at each point should not really be a continuum $S^1$.  Note that $M$ itself should probably be noncommutative also due to Planck scale effects, but we are not considering physical spacetime near the Planck scale, only the fibre. 

Such a value of $h$ of course means that the induced masses from the curvature of the fuzzy sphere are Planckian,
\begin{equation}\label{massh} m=\sqrt{\frac{l(l+1)}{h}}=\sqrt{\frac{l(l+1)\alpha_W}{4G}}\approx 0.09\sqrt{l(l+1)}m_p\end{equation}
and hence $l>0$ fields would be totally unreachable at particle accelerator energies. This is usually considered as a good thing as we do not see an infinite tower of massive fields at energies that we can observe. However, the possibility to take a reduced fuzzy sphere where $0\le l < 2j$ for small $j$ or more generally some finite QRG fibre (meaning the coordinate algebra for the fibre is finite-dimensional) offers a mechanism which, for the right model and right coupling strengths, might explain apparently finite families of particles with different masses, as discussed in\cite{ArgMa4,LiuMa2}. For example, $j=1/2$ for a reduced fuzzy sphere would mean a massless $SU(2)$ internal spin 0 field and a massive internal spin 1 field. The latter could appear not as a main field of the Standard Model but in some extended theories, such as the scalar field $\Delta$ in the Type II see-saw mechanism\cite{MagWet} which has weak $SU(2)$ spin 1. This, moreover, needs to have a very high mass, albeit still some orders below the Planck mass, see for example\cite{Gho}. The original (fermionic) seesaw  mechanism does have a Planckian mass right handed neutrino field\cite{Sch}, but this is, however, a singlet under weak $SU(2)$.

At present, we are only looking for a proof of concept rather than claiming that the fuzzy sphere lands us on observed particle spectra. In this regard, we would also need to obtain fermionic or spinor fields, some of which are weak $SU(2)$ doublets in the Standard Model or could be expected to be doublets under $SU(2)$ in other models. These could arise from spinors on the product spacetime rather than scalars as considered in the present paper. This is a more complicated project which will be pursed elsewhere but could potentially link up with ideas in \cite{Con,Cham} etc. coming from a different spectral triples approach. The latter Connes approach remains topical, e.g. \cite{DabSit}, albeit an issue there is what should be a Lorentzian spectral triple (now on the product). Some ideas for Lorentzian spectral triples are in \cite{Dev,PasSit}, for example.  In this respect, our more constructive approach starting with (\ref{prodmetric}) could have an advantage. For part of the QRG realisation, we can use the Dirac operator on the fuzzy sphere\cite{LirMa2}, where spinors are sections of charge $\pm1$ monopole bundles on the fuzzy sphere rather than functions as in the present paper. This in turn would lead to half-integer internal spin $l$ fields on spacetime $M$, e.g. $SU(2)$ doublets for $l=1/2$.

We can also look to upgrade a current global $SU(2)$ symmetry (or more likely approximate symmetry) of the Standard Model to a gauge symmetry, and a natural candidate here would be an $SU(2)_f$ `flavourspin' symmetry of some kind. This has been proposed in \cite{BerHer} within the more general paradigm of the Minimal Flavour Violation (MVF) ansatz. Here,  the three generations of fermions, which are identical other than in mass, are now components of a spin 1 vector representation of such an $SU(2)_f$, i.e. mixing the 3 generations of elementary particles `horizontally'. The various Yukawa couplings in this context are  $M_3(\R)$-valued, i.e.,  $Y_{ij}$ with two generation indices, and enter into the Lagrangian as terms of the form
\[   \CL_{\rm Yuk}=-\overline{Q_L}Y_u U_R\cdot \tilde H- \overline{Q_L}Y_d D_R \cdot H - \overline{L_L}Y_l E_R\cdot H + h.c. \]
To account for neutrino masses one has to go beyond the Standard Model, for example with Dirac spinor neutrinos acquiring mass via 
\[  \CL_\nu=-\overline{L_L}Y_\nu N_R\cdot\tilde H+ h.c.\]
The various fermion fields here are grouped in threes due to a generation index $i$. Thus, $L_L$ denotes the left-handed leptons, $Q_L$ left-handed quarks, both doublets under the weak $SU(2)$, while $U_R,D_R,E_R,N_R$ are right-handed up, down, electron and neutrino fields respectively, which are singlets under the weak $SU(2)$, but all these fields are vectors under $SU(2)_f$. (For example, $U_R$ contains right handed up, charm and top quarks as the triplet.) In the Type I seesaw mechanism for neutrinos, one would take instead Majorana spinors and add a mass term $M\overline{N_R}N_R^c$, where $M$  is a Planck order mass for $N_R$. Moreover, following Cabbibo\cite{Cab}, one can upgrade the Yukawa couplings above to dynamical fields $Y_u,Y_d,Y_l, Y_\nu$ and, for the Type II seesaw mechanism, a field $Y_\Delta$, all with 2 generation indices $i,j$. These decompose as `flavourspin' $l=0,1,2$ but are otherwise real scalar fields. This is exactly what we would obtain from a reduced fuzzy sphere at $j=1$ as a real scaler field on the product. The $l=0$ component is flavourless and not discussed in \cite{BerHer} but some distinctive properties of the $l=1,2$ fields are discussed from a phenomenological point of view. If so, then by our results the different spin components of each type of Yukawa field could appear with masses in the ratios $0,1,\sqrt{3}$ according to the values of $\sqrt{l(l+1)}$ if the real scalar field on the product is massless.

Finally, we note that the associated $SU(2)$ gauge field $\bar A^i_\mu$ in this new scenario amounts to an extremely weak new force. For the sake of discussion here, we reverse our calculation (\ref{massh}) for the generated mass $m$ to land on a particle accelerator type mass 1 TeV or 1.8 $\times 10^{-24}$kg, say. Then the `flavour' fine-structure constant that we need is (in our units as above)
\[ \alpha_f= {4 G m^2 \over l(l+1)}={4\over l(l+1)}\left({m\over m_P}\right)^2 \approx 1.3 \times 10^{-32}\]
if we set $l=1$. This is of the same order as the effective gravitational fine structure constant for particles of this same mass, given by $\alpha_G=G m^2$ for two particles of mass $m$, see e.g.\cite{Jen}.  In other words, for this application, the hypothetical $SU(2)$ gauge symmetry coming out of the cross terms in the product metric has a similar strength as gravity on the spacetime factor for the mass generated, which seems appropriate.

 \end{document}